\documentclass[a4paper,12pt,halfparskip]{scrartcl}
\pdfoutput=1
\usepackage[utf8]{inputenc}
\usepackage[english]{babel}
\usepackage{cite}
\usepackage{amsmath}
\usepackage{amssymb}
\usepackage{dsfont}
\usepackage{graphicx}
\usepackage{hyperref}
\usepackage{setspace}
\usepackage[font=small,labelfont=bf,textfont=it,margin=10pt,format=plain,indention=10pt]{caption}
\numberwithin{equation}{section}
\usepackage[automark,headsepline,ilines]{scrpage2}
\usepackage{nicefrac}
\usepackage{authblk}
\usepackage{subfig}
\usepackage{longtable}
\usepackage{lscape}
\usepackage{sidecap}
\hypersetup{}

\titlehead{\hfill{}WUB/13-01}
\title{A recursive approach to determine correlation functions in multibaryon systems}

\author{Jana Günther,\,\, Bálint C. Tóth\,\, and\,\, Lukas Varnhorst}
\date{May 28, 2013}
\affil{Department of Physics, Bergische Universität Wuppertal, D-42119 Wuppertal, Germany}

\begin{document}

\maketitle

\begin{abstract}
We propose a recursive algorithm for the calculation of multibaryon
correlation functions that combines the advantages of a recursive approach
with those of the recently proposed unified contraction algorithm. The
independent components of the correlators are built recursively by adding the
baryons one after the other in a given order. The list of nonzero independent
components is also constructed in a recursive manner, significantly reducing
the resources required for this step. We computed the number of operations
required to calculate the correlators up to $^8 \text{Be}$, and observed a
significant speedup compared to other techniques. For the calculation of
$^4\text{He}$ and $^8\text{Be}$ correlation functions in the fully
relativistic case $\mathcal O(10^8)$ operations are required, whereas for
nonrelativistic operators this number can be reduced to e.g. $\mathcal
O(10^4)$ in the case of $^4\text{He}$.
\end{abstract}

\section{Introduction}
\label{sec:intro}

Quantum Chromodynamics (QCD), the theory of the strong interaction was first
introduced to describe the strong nuclear binding forces. Given this fact QCD
is expected to be able to predict the masses and properties of atomic
nuclei. Due to the strong coupling at low energies nonperturbative techniques
such as lattice QCD (LQCD) are required to study bound states in QCD. In
principle the tools for such studies are at hand and several calculations in
order to examine light nuclei \cite{Yamazaki:2009ua,Beane:2012vq,
  Yamazaki:2012hi} and the nuclear force
\cite{Ishii:2006ec,Aoki:2009ji,Doi:2011gq} have been performed recently.
However, the enormous amount of Wick contractions necessary for the evaluation
of correlation functions of atomic nuclei is a severe problem in such
calculations.

The number of Wick contractions for the na\"ive evaluation of correlation
functions of multibaryon systems scales as $n_u!\,n_d!\,n_s!$, where $n_u$,
$n_d$ and $n_s$ are the number of $u$, $d$ and $s$ quarks in the system,
respectively. Furthermore, for each Wick contraction one has to evaluate the
sum over all color and spin indices. These sums scale exponentially with the
number of baryons in the system. As a consequence, the total number of
required operations scales as $n_u!\,n_d!\,n_s!\,6^A\,4^A$, where $A$ is the
atomic mass number. The introduction of more complicated spatial baryonic
wave functions will increase this number further.

For the related, but somewhat simpler case, where the system consists of a
large number of mesons, several efficient methods exist
\cite{Detmold:2010au,Detmold:2012wc}. The most recent of these techniques
allows for the study of systems containing up to 72 pions
\cite{Detmold:2012wc}.

For systems comprised of baryons there has also been substantial progress
recently in reducing this computational challenge. In
Ref.\ \cite{Yamazaki:2009ua} the number of contractions has been reduced
significantly by exploiting the permutation symmetry of the quark operators. A
further improvement has been achieved in \cite{Doi:2012xd}, where the combined
permutations of color and spin indices are used to create a unified list of
independent contractions. While this method reduces the amount of contractions
to be evaluated on each gauge configuration significantly, the creation of the
list of independent contractions remains difficult. This is due to the fact
that the full set of possible contractions, which scale factorially and
exponentially in the number of quarks, has to be applied once to determine the
coefficients in the list. For small systems it is possible to carry out this
calculation once, but it becomes quickly impractical for larger systems.  The
method proposed in Ref.\ \cite{Detmold:2012eu} besides being able to consider
multiple source locations, brings an improvement by generating the list of
terms to be contracted recursively. The determinant algorithm of
Ref.\ \cite{Detmold:2012eu} can further reduce the computational cost in the
case of certain large nuclei by transforming the factorially scaling task of
calculating Wick contractions into the polynomially scaling task of
calculating determinants.

The purpose of this paper is to propose an efficient method for the
calculation of baryonic correlation functions of the form
\begin{equation}
 [C^{(N)}]_{\alpha_1 \alpha_2 \ldots \alpha_{N}}^{\alpha_1' \alpha_2' \ldots
   \alpha_{N}'}(\vec x_1, \vec x_2, \ldots, \vec x_{N}, t) = \left<
 \prod_{k=1}^{N} B_{\alpha_k}(\vec x_k, t) \, \prod_{l=1}^{N} \overline
 B^{\alpha_l'}(\vec 0, 0) \right>
 \label{eqn:C_baryons}
\end{equation}
that combines the advantages of the recursive approach with those of the
algorithm introduced in \cite{Doi:2012xd}. The interpolating baryon operators
are of the form
\begin{subequations}
\begin{align}
 B_{\alpha} &= \varepsilon_{abc} \, (\Gamma_1)_{\alpha\beta} (q_1)_{\beta;a}
 \, [ (q_2)_{\gamma;b} (\Gamma_2)_{\gamma\delta} (q_3)_{\delta;c} ], \\
 \overline B^{\alpha} &= \varepsilon^{abc} \, (\Gamma_1)^{\alpha\beta}
 (\overline q_1)^{\beta;a} \, [ (\overline q_2)^{\gamma;b}
   (\Gamma_2)^{\gamma\delta} (\overline q_3)^{\delta;c} ],
\end{align}
\end{subequations}
where the quark operators $q_n \in \{u,d,s\}$ and $ \overline q_n \in
\{\overline u,\overline d,\overline s\}$ are all taken at the same spacetime
point. Here and throughout in the paper Latin indices correspond to color
degrees of freedom (DoFs) while Greek indices correspond to the spin DoFs
associated with the quark operators. For notational convenience all upper
indices correspond to quark operators at the source while lower indices
correspond to quark operators at the sink. The choice of $\Gamma_1 = \mathds
1$ and $\Gamma_2 = C\gamma_5$ yields the interpolating operators for the
proton with $(q_1, q_2, q_3) = (u, u, d)$ and for the neutron with $(q_1, q_2,
q_3) = (d,u,d)$.

The paper is organized as follows. First we review the unified contraction
algorithm in Section \ref{sec:UCA}. In Section \ref{sec:antisym} a method is
introduced to construct antisymmetric tensors out of small building blocks in
a recursive way. This procedure is applied in Section \ref{sec:one_source} to
construct correlation functions of multibaryon systems with one quark source
and one baryon sink. In Section \ref{sec:spin_proj} a method is described to
reduce the number of necessary operations when only the projection of the
correlation function to a certain angular momentum state is of interest. This
is followed by the generalization of the method to an arbitrary number of
quark sources and baryon sinks in Section \ref{sec:many_sources}, which allows
the calculation of arbitrarily complex correlation functions.  The case of
atomic nuclei is discussed in detail in Section \ref{sec:nuclei}. Finally,
after comparing the efficiency of our method with that of other recent
algorithms in Section \ref{sec:efficiency} we conclude in Section
\ref{sec:conclusions}.

\section{The unified contraction algorithm}
\label{sec:UCA}

To provide a self-contained presentation, we briefly review in this section
the unified contraction algorithm introduced in Ref.\ \cite{Doi:2012xd}. For
the construction of multibaryon correlation functions it is useful to define
blocks of quark propagators which correspond to the contractions of three
quarks at the source with a baryon at the sink.  This blocking procedure,
which was successfully used both for the study of light nuclei
\cite{Yamazaki:2009ua, Beane:2012vq, Yamazaki:2012hi} and for the study of
several-nucleon forces \cite{Ishii:2006ec,Aoki:2009ji,Doi:2011gq}, has several
advantages.  First it already reduces the number of contractions to
evaluate. Second it allows to carry out the projection of individual baryons
to definite momentum or to introduce different baryon sinks prior to the
expensive calculation of the correlation function.  The blocks are generally
defined as
\begin{equation}
 f_B^{q_1, q_2, q_3}(t, \delta; \alpha, \beta, \gamma; a, b, c) = \sum_{\vec
   x}s(\vec x) \left< B_\delta(\vec x, t) \cdot \overline q_1^{\alpha;a} \overline q_2^{\beta;b}
 \overline q_3^{\gamma, c} \right>.
\end{equation}
Here $\delta$ is the spin index of the baryon $B$; $\alpha$, $\beta$ and
$\gamma$ are the spin indices of the three quarks $\overline q_1$, $\overline
q_2$ and $\overline q_3$; and $a$, $b$, $c$ are the corresponding color
indices. The forms of all three quark source operators are taken to be the
same.  The function $s(\vec x)$ characterizes the form of the baryon sink. A
common choice is the projection to zero momentum $s(\vec x)\propto 1$, which
is often needed e.g. when the mass of a bound state is to be extracted from a
correlation function.  For the moment it is assumed that the sink function is
the same for all baryons and the case of different sinks is discussed later.
For notational convenience the 4 spinor and 3 color degrees of freedom
associated with a quark of a given flavor can be combined to form spinor-color
indices $\xi^{(q)}$, which can take the values $1,2,\ldots,12$. Using these
combined indices a block can be written as
\begin{equation}
 f_B^{q_1,q_2,q_3}(t, \delta; \xi^{(q_1)}, \xi^{(q_2)}, \xi^{(q_3)}) =
 \sum_{\vec x}s(\vec x) \left< B_\delta(\vec x, t) \cdot \overline q_1^{\xi^{(q_1)}}
 \overline q_2^{\xi^{(q_2)}} \overline q_3^{\xi^{(q_3)}} \right>.
\end{equation}
In the case of a system consisting of protons $q_1 = q_2 = u$, therefore,
the above tensor is antisymmetric in the indices $\xi^{(q_1)}$ and
$\xi^{(q_2)}$.

Using the above defined blocks the correlation function of $N$ baryons can be
expressed as\footnote{We use the Einstein summation convention throughout the
  paper, that is, over each index appearing twice the sum is automatically
  understood.}
\begin{multline}
[C^{(N)}]_{\delta_1, \delta_2, \ldots, \delta_{N}}^{\alpha_1, \alpha_2,
  \ldots, \alpha_{N}}(t) = \\ \sum_{\sigma\in\Sigma} f_{B_1}^{q_1,q_2,q_3}(t,
\delta_1; \xi_1^{(q_1)}, \xi_2^{(q_2)}, \xi_3^{(q_3)}) \ldots
f_{B_{N}}^{q_1,q_2,q_3}(t, \delta_{N}; \xi_{3N-2}^{(q_1)}, \xi_{3N-1}^{(q_2)},
\xi_{3N}^{(q_3)}) \\ \cdot G^{B_1}(\alpha_1; \xi_{\sigma(1)}^{(q_1)},
\xi_{\sigma(2)}^{(q_2)}, \xi_{\sigma(3)}^{(q_3)}) \ldots
G^{B_{N}}(\alpha_{N};\xi_{\sigma(3N-2)}^{(q_1)}, \xi_{\sigma(3N-1)}^{(q_2)},
\xi_{\sigma(3N)}^{(q_3)}) \, \operatorname{sgn}(\sigma),
\end{multline}
where the objects $G_B$ are combinations of $\Gamma$-matrices and
$\varepsilon$-tensors suitable for a baryon $B$:
\begin{equation}
 G^{B}(\alpha; \xi^{(q_1)}, \xi^{(q_2)}, \xi^{(q_3)}) :=
 (\Gamma_1)^{\alpha\beta(\xi^{(q_1)})}
 (\Gamma_2)^{\beta(\xi^{(q_2)})\beta(\xi^{(q_3)})}
 \varepsilon^{c(\xi^{(q_1)})c(\xi^{(q_2)})c(\xi^{(q_3)})}.
 \label{eqn:def_g}
\end{equation}
Here $\beta(\xi)$ is the spin-index part of $\xi$ and $c(\xi)$ is the
color-index part and $\Sigma$ is the set of all permutations that permute the
indices associated with the different quark flavors $q_k$ separately. The
product of blocks $f_B^{q_1,q_2,q_3}$ does not depend on the permutations
$\sigma$ and hence the correlation function can be written in the form
\begin{multline}
 [C^{(N)}]_{\delta_1, \delta_2, \ldots, \delta_{N}}^{\alpha_1, \alpha_2,
   \ldots, \alpha_{N}}(t) = f_{B_1}^{q_1,q_2,q_3}(t, \delta_1; \xi_1^{(q_1)},
 \xi_2^{(q_2)}, \xi_3^{(q_3)}) \ldots f_{B_{N}}^{q_1,q_2,q_3}(t, \delta_{N};
 \xi_{3N-2}^{(q_1)}, \xi_{3N-1}^{(q_2)}, \xi_{3N}^{(q_3)}) \\ \cdot
 L^{(N)}(\alpha_1,\ldots,\alpha_{N};\xi_{1}^{(q_1)}, \xi_{2}^{(q_2)},
 \xi_{3}^{(q_3)}, \ldots, \xi_{3N-2}^{(q_1)}, \xi_{3N-1}^{(q_2)},
 \xi_{3N}^{(q_3)})
 \label{eqn:f_L_contr}
\end{multline}
with the tensor
\begin{multline}
 L^{(N)}(\alpha_1,\ldots,\alpha_{N};\xi_{1}^{(q_1)}, \xi_{2}^{(q_2)},
 \xi_{3}^{(q_3)}, \ldots, \xi_{3N-2}^{(q_1)}, \xi_{3N-1}^{(q_2)},
 \xi_{3N}^{(q_3)}) = \\ \sum_{\sigma\in\Sigma} G^{B_1}(\alpha_1;
 \xi_{\sigma(1)}^{(q_1)}, \xi_{\sigma(2)}^{(q_2)}, \xi_{\sigma(3)}^{(q_3)})
 \ldots G^{B_{N}}(\alpha_{N}; \xi_{\sigma(3N-2)}^{(q_1)},
 \xi_{\sigma(3N-1)}^{(q_2)}, \xi_{\sigma(3N)}^{(q_3)}) \,
 \operatorname{sgn}(\sigma).
 \label{eqn:tensor_L}
\end{multline}
In the unified contraction algorithm the object $L$ is generated by explicitly
performing all permutations according to eqn.\ (\ref{eqn:tensor_L}). Since the
$G_B$'s are very sparsely populated tensors in most cases $L$ is also
sparse. It is then proposed to consider only those components of the product
$f_{B_1}^{q_1,q_2,q_3}\ldots f_{B_N}^{q_1,q_2,q_3}$ which are contracted with
the nonzero components of $L$.

\section{Recursive construction of antisymmetric tensors}
\label{sec:antisym}

The object $L$ has a high degree of symmetry which reduces the number of its
independent components. From the definition it is straightforward to see that
$L$ is antisymmetric under the exchange of two indices $\xi$ as long as they
belong to the same quark flavor. It also possesses a number of spin indices
$\alpha_1,\ldots,\alpha_{N}$ corresponding to baryons of types $B_1, B_2,
\ldots, B_{N}$. From the Pauli principle it follows that the correlator
$[C^{(N)}]_{\delta_1, \delta_2, \ldots, \delta_{N}}^{\alpha_1, \alpha_2,
  \ldots, \alpha_{N}}(t)$ has to be antisymmetric under the exchange of any two
indices $\alpha$ corresponding to the same type of baryon. Hence the same
property has to hold for $L$.

It can be shown using only this antisymmetric property that the maximal number
of independent components e.g. in the case of $^4 \text{He}$ is $30735936$,
whereas in the case of $^8 \text{Be}$ it is $1$. Since the objects $G$ are
very sparse, many of these components are expected to be zero.

A component of a tensor $X(\xi_1, \xi_2, \ldots, \xi_l)$, which is
antisymmetric in the indices $\xi_1, \xi_2, \ldots, \xi_l$, each ranging from
$1$ to $k$, can be uniquely defined by a $k$-tuple $\boldsymbol A\{\xi\} =
(n(1), n(2), \ldots ,n(k))$, where $n(i)$ denotes how often the value $i$
occurs amongst the $l$ indices in the set $\{\xi\}=\{\xi_1, \xi_2, \ldots,
\xi_l\}$. As a consequence of the antisymmetry all components with $n(i)>1$
vanish. The component associated with such a tuple is the component where all
values $i$ for which $n(i)=1$ occur amongst the indices in ascending order.
All other components can be constructed using permutations of the indices and
taking the sign of the permutation into account. For example, if $X$ is a
tensor with three antisymmetric indices, each ranging from one to four, the
tuple $\boldsymbol A\{\xi\} = (1,0,1,1)$ corresponds to the component
$X(1,3,4)$. If a tensor is antisymmetric in several groups of indices
independently, then several independent tuples can be defined, one for each
group of indices.

If $X$ is an antisymmetric tensor with $k$ indices and $Y$ is an antisymmetric
tensor with $l$ indices, then their antisymmetric product $Z = X \bullet Y$ is
a tensor with $k+l$ antisymmetric indices, whose components are defined
as\footnote{In the definition of this product the normalization factors have
  been removed deliberately to speed up the computation. These factors will be
  reintroduced when the correlation function is calculated.}
\begin{equation}
 (X \bullet Y)(\boldsymbol z) := Z(\boldsymbol z) = \sum_{\boldsymbol z =
    \boldsymbol x + \boldsymbol y} X(\boldsymbol x) Y(\boldsymbol y)
  \operatorname{sgn}(\boldsymbol x| \boldsymbol y),
\end{equation}
where the tuples
\begin{align}
 \boldsymbol z &= \boldsymbol A\{\xi_1, \ldots, \xi_{k+l}\} \\
 \boldsymbol x &= \boldsymbol A\{\xi_1, \ldots, \xi_k\} \\
 \boldsymbol y &= \boldsymbol A\{\xi_{k+1},\ldots,\xi_{k+l}\}
\end{align}
identify the antisymmetric components and
\begin{equation}
\operatorname{sgn} (\boldsymbol x|\boldsymbol y) =                                                                                                     
\prod_{\substack{i>j \\ y_j=1}} (-1)^{x_i}
\end{equation}
is the sign of the permutation that is necessary to bring
the indices of the tensors $X$ and $Y$ into ascending order.

If each tensor has $r$ independent groups of antisymmetric indices then
each such group is described by an individual tuple. In this case the
antisymmetrized product can be written as
\begin{multline}
 (X \bullet Y)(\boldsymbol z_1, \boldsymbol z_2, \ldots, \boldsymbol z_r) :=
  \\ Z(\boldsymbol z_1, \boldsymbol z_2, \ldots, \boldsymbol z_r) =
  \sum_{\substack{\boldsymbol z_1 = \boldsymbol x_1 + \boldsymbol y_1
      \\ \boldsymbol z_2 = \boldsymbol x_2 + \boldsymbol y_2 \\ \cdots
      \\ \boldsymbol z_r = \boldsymbol x_r + \boldsymbol y_r}} X(\boldsymbol
  x_1, \boldsymbol x_2, \ldots, \boldsymbol x_r)Y(\boldsymbol y_1, \boldsymbol
  y_2, \ldots, \boldsymbol y_r) \times \\ \times
  \operatorname{sgn}(\boldsymbol x_1| \boldsymbol
  y_1)\operatorname{sgn}(\boldsymbol x_2| \boldsymbol
  y_2)\ldots\operatorname{sgn}(\boldsymbol x_r| \boldsymbol y_r).
 \label{eqn:antis_product_2}
\end{multline}
In the following it will be often required to antisymmetrize only the subset
of the quark spinor-color indices $\xi$ that corresponds to a given quark flavor
$q$. In this case it will be useful to write $\mathbf A^{(q)}\{\xi_1, \xi_2,
\ldots, \xi_n\}$ for the tuple of indices associated with the respective quark
flavor. In the case of the spinor indices $\alpha$ and $\delta$ of the baryons
a similar notation is adopted: here $\mathbf A^{(B)}\{\alpha_1, \alpha_2
\ldots \alpha_n \}$ is the tuple associated with the antisymmetrization of
only those spinor indices that correspond to the baryon type $B$.

In the special case where an antisymmetric tensor can be written as $X^{(n)}
= Y_1 \bullet Y_2 \bullet \cdots \bullet Y_n$, a recursion relation $X^{(i)} =
X^{(i-1)} \bullet Y_i$ can be set up with the starting condition $X^{(1)} =
Y_1$. The usage of this recursion relation is often much more efficient than
the direct evaluation of the product $Y_1 \bullet Y_2 \bullet \cdots \bullet
Y_n$. Assuming that there are $r$ groups of antisymmetric indices, each
index in the $p$-th group can take values from 1 to $m_p$, and in the
$p$-th group at the stage $X^{(i)}$ there are $n_p$ indices, then the number of
components at the intermediate step is
\begin{subequations}
\begin{equation}
 P^{(i)} = \prod_{p=1}^r C(n_p, m_p-n_p),
 \label{eqn:gen_P}
\end{equation}
where $C(n_1,n_2,\ldots,n_m) = (n_1+n_2+\ldots+n_m)! / n_1! n_2! \ldots n_m!$
are the multinomial coefficients. The number of operations required to go from
$X^{(i)}$ to $X^{(i+1)}$ is
\begin{equation}
 Q^{(i)} = \prod_{p=1}^r C(n_p, l_p, m_p-n_p-l_p),
 \label{eqn:gen_Q}
\end{equation}
\end{subequations}
where $l_p$ is the number of indices in the $p$-th index group of $Y_{i+1}$.

\begin{SCfigure}
 \begin{centering}
  \includegraphics[width=0.5\textwidth]{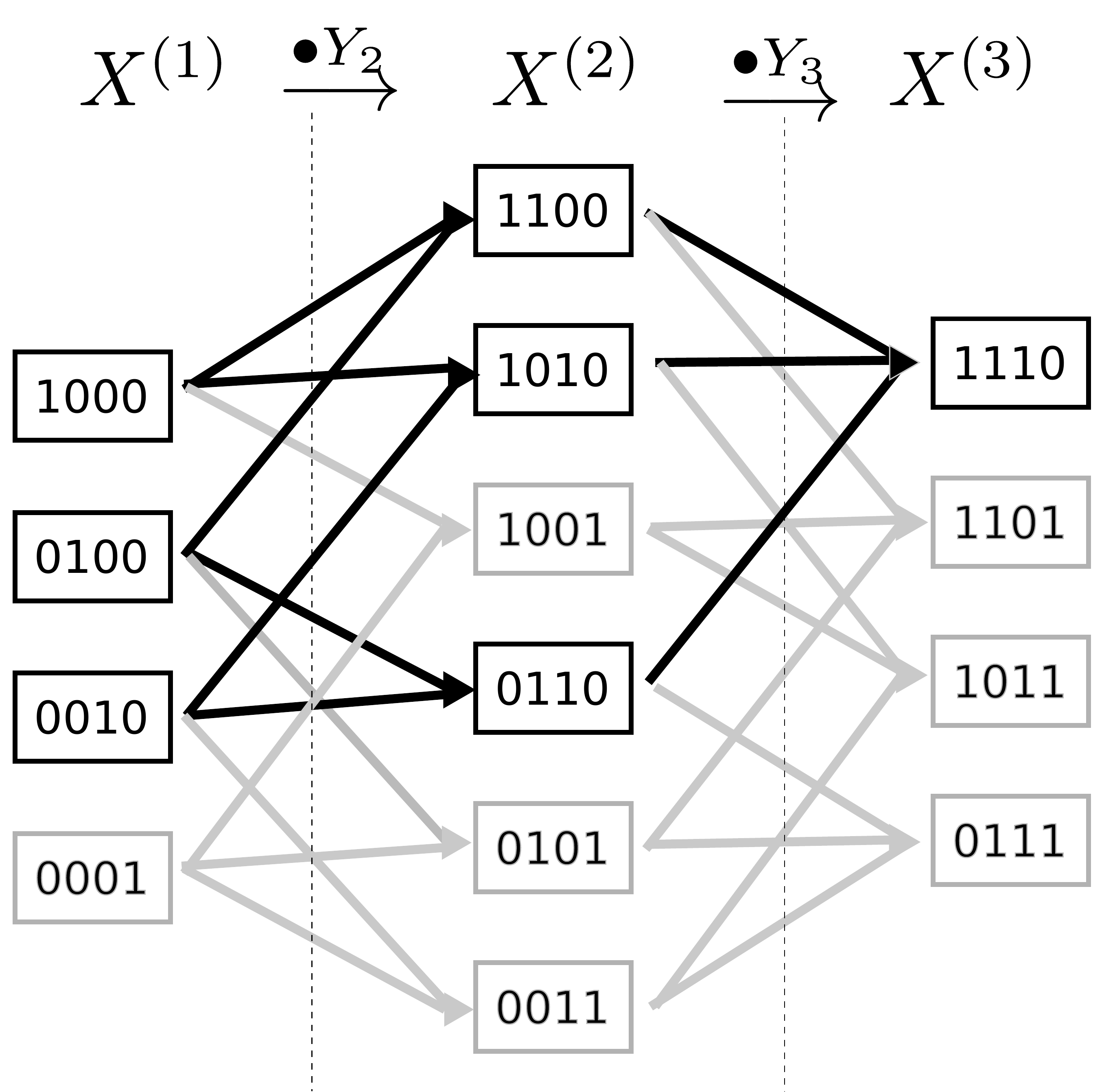}
  \caption[The computational savings of the recursive algorithm in the case of
    a simle example.]{The computational savings of the recursive algorithm in
    the case when only one antisymmetric component of the result $X^{(3)}$ is
    of interest. $X^{(3)}$ has one group of antisymmetric indices, and each
    $Y_{i}$ possesses only one index.  The boxes in the three columns
    represent the antisymmetric components of the respective stages $X^{(1)}$,
    $X^{(2)}$ and $X^{(3)}$. The number in each box corresponds to the tuple
    associated with the antisymmetric component. The arrows show which
    components are used to construct the components of the next stage. If the
    gray boxes in the tensor $X^{(3)}$ are not needed, then the gray boxes in
    all other tensors are also not needed. Only the solid black operations
    have to be performed and the gray operations can be omitted.
  \label{fig:antisym_savings}}
 \end{centering}
\end{SCfigure}

A further reduction of the computational effort can be achieved when not all
components of $X^{(n)}$ at the final stage are of interest. When evaluating the
product $X^{(n)}=X^{(n-1)}\bullet Y_n$ only those summands have to be
considered which contribute to a component of interest in $X^{(n)}$. Using
this property not only the computational effort of evaluating this particular
product can be reduced, but also some of the components of $X^{(n-1)}$ may not
be required for the evaluation of the product at all. Therefore, these
components of $X^{(n-1)}$ are not of interest and need not be computed in the
previous recursion step. This argumentation can then be repeated for all
recursion steps and often leads to a significant reduction of computational
effort. The procedure is demonstrated in Figure \ref{fig:antisym_savings} for the
case where $X^{(n)}$ has only one group of antisymmetric indices,  each
index ranging from 1 to 4. The tensors $Y_i$ here possess only one index of
the same format.  If it is assumed that at the final stage only the
black component of $X^{(3)}$ is of interest then a significant reduction in the
computational effort is achieved.

If it is known in advance which components of $X^{(n)}$ are of interest, then
it is useful to determine which components of $X^{(n-1)}$ are required for
their computation and to make a list $\Lambda^{(n-1)}$ of the required
operations.  This procedure can be repeated successively for all previous
intermediate steps until $X^{(1)}$ is reached. For the computation of the
desired components of $X^{(n)}$ one then has to perform only the operations
contained in the lists $\Lambda^{(1)}, \Lambda^{(2)}, \ldots,
\Lambda^{(n-1)}$.

In practice the lists $\Lambda^{(i)}$ are usually much larger than the tensors
$X^{(i)}$. Therefore, the memory requirement of the resulting algorithm is
dominated by these lists. This problem can be circumvented by storing at each
stage $i$ only the list of the required components of $X^{(i)}$. Then the
reconstruction of the list of operations $\Lambda^{(i-1)}$ introduces a
relatively small overhead, but the amount of memory used by the algorithm is
reduced significantly.

\section{Correlation functions with one quark source/sink}
\label{sec:one_source}

Using the notation from the previous section the tensor $L$ can be written
in the form\footnote{The generalization to systems with additional quark flavors is
  straightforward. For notational convenience we restrict ourselves here to only
  $u$, $d$ and $s$ quarks.}
\begin{equation}
 L(\boldsymbol A^{(B_a)}\{\alpha\}, \boldsymbol A^{(B_b)}\{\alpha\}, \ldots,
 \boldsymbol A^{(u)}\{\xi\}, \boldsymbol A^{(d)}\{\xi\}, \boldsymbol
 A^{(s)}\{\xi\}).
\end{equation}
Here $B_a, B_b, \ldots$ are the different types of baryons in the
system. Using a similar argumentation the object $G^B$ can be written as
\begin{equation}
 G^B(\alpha, \boldsymbol A^{(u)}\{\xi\}, \boldsymbol A^{(d)}\{\xi\}, \boldsymbol
 A^{(s)}\{\xi\}).
\end{equation}
Although the index $\alpha$ is just a single spin index, it can be expressed
through 4-tuples $\boldsymbol A^{(B_a)}\{\alpha\}, \boldsymbol
A^{(B_b)}\{\alpha\}, \ldots$ to bring $G^B$ into the same form as $L$. Out of
these tuples only the one corresponding to the baryon $B$ will have a single
nonzero entry.

The tensors $L^{(n)}$ defined in eqn.\ (\ref{eqn:tensor_L}) corresponding to
$n$ baryons fulfill the recursion relation
\begin{equation}
 L^{(n+1)} = L^{(n)} \bullet G_{B_{n+1}}
 \label{eqn:recr_L}
\end{equation}
with the starting condition
\begin{equation}
 L^{(1)} = G_{B_1}.
\end{equation}
Here ``$\bullet$'' denotes the antisymmetric product with multiple groups of
antisymmetric indices as defined in equation (\ref{eqn:antis_product_2}).  In
general $G_{B_{n}}$ can be a different tensor describing a different type of
baryon for each $n$. The objects $G_{B_{i}}$ are products of
$\varepsilon$-tensors and $\Gamma$-matrices and are therefore often
sparse. Hence the evaluation of the above recursion can be done very
efficiently if only the nonzero components are stored.

According to eqn.\ (\ref{eqn:f_L_contr}) the correlation function for a given
gauge configuration is obtained by evaluating the contraction of $L$ with the
product
\begin{multline}
 F^{(N)}(\delta_1,\ldots, \delta_{N}; t; \xi_1^{(q_1)}, \xi_2^{(q_2)},
 \xi_3^{(q_3)},\ldots, \xi_{3N-2}^{(q_1)}, \xi_{3N-1}^{(q_2)},
 \xi_{3N}^{(q_3)}) \\ := f_{B_1}^{q_1,q_2,q_3}(t, \delta_1; \xi_1^{(q_1)},
 \xi_2^{(q_2)}, \xi_3^{(q_3)}) \ldots f_{B_{N}}^{q_1,q_2,q_3}(t, \delta_{N};
 \xi_{3N-2}^{(q_1)}, \xi_{3N-1}^{(q_2)}, \xi_{3N}^{(q_3)}).
\end{multline}
To do so one could in principle construct all the components of $L$
explicitly.  However, it is computationally more efficient to exploit the
antisymmetry of $L$ directly: The tensor $F$ can be projected to a tensor
$F_{-}$ which is antisymmetric in all indices corresponding to the same quark
flavor or the same baryon.  Only this antisymmetric projection contributes to
the correlation function, and therefore, only the contraction between $L$ and
$F_{-}$ has to be evaluated.

$F_{-}$ possesses the same antisymmetry structure as $L$, thus it can be
written in the form
\begin{equation}
 F^{(n)}_{-}(\boldsymbol A^{(B_a)}\{\delta\}, \boldsymbol A^{(B_b)}\{\delta\},
 \ldots, \boldsymbol A^{(u)}\{\xi\}, \boldsymbol A^{(d)}\{\xi\},
 \boldsymbol A^{(s)}\{\xi\}).
\end{equation}
Since $F_{-}$ is composed of the independent factors $f_{B}^{q_1,q_2,q_3}$, a
similar recursion relation
\begin{equation}
 F_{-}^{(n+1)} = F_{-}^{(n)} \bullet f_{B_{n+1}}^{q_1,q_2,q_3} \label{eqn:recr_F}
\end{equation}
with the starting condition $F_{-}^{(1)} = f_{B_{1}}^{q_1,q_2,q_3}$ can be
defined.  In order to apply the above recursion relation one has to calculate
explicitly the independent antisymmetric components of
$f_{B_{i}}^{q_1,q_2,q_3}$ by applying all possible permutations of the combined
spinor-color indices. Since there are only three such indices in each factor,
the computational effort associated with these antisymmetrizations can be
neglected compared to the evaluation of the recursion steps of
(\ref{eqn:recr_F}).

Once both $F_{-}$ and $L$ are ready, the correlation function can be extracted
by performing the contraction
\begin{multline}
C^{(N)}(t; \boldsymbol A^{(B_a)}\{\delta\}, \ldots, \boldsymbol
A^{(B_a)}\{\alpha\}, \ldots) \\ 
= \frac{1}{\mathcal N} \sum_{\substack{\boldsymbol A^{(q_i)}\{\xi\} \\
i\in\{a,b,c\}}} F^{(N)}_{-}(\boldsymbol A^{(B_a)}\{\delta\}, \ldots,
\boldsymbol A^{(u)}\{\xi\}, \ldots) \cdot L^{(N)}(\boldsymbol
A^{(B_a)}\{\alpha\}, \ldots, \boldsymbol A^{(u)}\{\xi\},\ldots)
 \label{eqn:C_final}
\end{multline}
with the normalization factor
\begin{equation}
 \mathcal N = n_{q_a}!\,n_{q_b}!\ldots(n_{B_a}!\,n_{B_b}!\ldots)^2,
\end{equation}
where $n_{q_i}$ is the number of quarks of flavor $q_i$ and $n_{B_i}$ is the
number of baryons of type $B_i$ in the system.

The final result $C^{(N)}(\boldsymbol A^{(B_a)}\{\delta\}, \ldots, \boldsymbol
A^{(B_a)}\{\alpha\}, \ldots)$ itself is an antisymmetric tensor. All components
can be reconstructed by taking into account the respective permutations to
change the ordering of the indices.

Due to the sparse nature of $G_{B_i}$ the computational cost of the
determination of $L^{(N)}$ is very low. Furthermore, this construction has to
be performed only once independent of any gauge configuration. The calculation
of $F_{-}^{(m)}, m=1,2,\ldots,N$ is computationally much more demanding and
has to be performed on each gauge configuration. The maximal number of
components at the intermediate stage $F_{-}^{(m)}$ is given by the formula
\begin{equation}
 P(n^{(m)}_{B_1}, n^{(m)}_{B_2}, \ldots) = \prod_i C(n^{(m)}_{B_i},
 4-n^{(m)}_{B_i}) \prod_j C(n^{(m)}_{q_j}, 12-n^{(m)}_{q_j}),
\label{eqn:rec_P}
\end{equation}
where $n^{(m)}_{B_i}$ is the number of baryons $B_i$ at the intermediate stage
$m$ and $n^{(m)}_{q_j}$ is the number of quarks of flavor $q_j$ at this stage.

In a similar way the maximal number of operations required for adding the
baryon $B_k$ in the recursion step $F_{-}^{(m)} \rightarrow F_{-}^{(m+1)}$ is
\begin{multline}
 Q_{B_k}(n^{(m)}_{B_1}, n^{(m)}_{B_2}, \ldots) = 
\prod_i 
\begin{cases}
C(n^{(m)}_{B_i}, 0, 4-n^{(m)}_{B_i}) & \text{for} \ i \neq k \\
C(n^{(m)}_{B_i}, 1, 4-1-n^{(m)}_{B_i}) & \text{for} \ i = k
\end{cases}\\
\times \prod_j 
\begin{cases}
C(n^{(m)}_{q_j}, 0, 12-n^{(m)}_{q_j}) & \text{for} \ j \neq k \\
C(n^{(m)}_{q_j}, N(q_j,B_k), 12-n^{(m)}_{q_j}-N(q_j,B_k)) & \text{for} \ j = k,
\end{cases}
\label{eqn:rec_Q}
\end{multline}
where $N({q_j},{B_k})$ is the number of quarks of flavor $q_j$ in the baryon
of type $B_k$.  Formulas (\ref{eqn:rec_P}) and (\ref{eqn:rec_Q}) are special
cases of the more general relations (\ref{eqn:gen_P}) and (\ref{eqn:gen_Q}).

The estimates $P$ and $Q$ obtained above are only upper bounds on the real
computational cost. Due to the large number of zero components of $L^{(N)}$
large savings in comparison with these numbers are possible. During the
recursive constructions of $F^{(1)}_{-}$, $F^{(2)}_{-}$, $\ldots$,
$F^{(N)}_{-}$ only those components have to be calculated, which in the end
contribute to a component contracted with a nonzero component of $L^{(N)}$.
To do so one calculates the lists $\Lambda^{(1)}, \Lambda^{(2)}, \ldots,
\Lambda^{(N-1)}$ using the procedure described in Section
\ref{sec:antisym}. Then only these operations have to be performed for each
gauge configuration.

\section{Projection to angular momentum states\label{angular_momentum}}
\label{sec:spin_proj}

The procedure described in the previous section allows the generation of all
spinor components of the correlation function. However, in many cases not all
spinor components are of interest, but the correlation function is to be
projected to a definite angular momentum state. It is often the case that not
all components contribute to this projection and hence one wants to avoid the
unnecessary computational effort.

Let $\mathcal M$ be an arbitrary tensor that projects the correlation function
to the required spin state such that
\begin{equation}
 C_{\mathcal M}(t) = \mathcal M_{\delta_1, \delta_2, \ldots,
   \delta_{N}}^{\alpha_1, \alpha_2, \ldots, \alpha_{N}} [C^{(N)}]_{\delta_1,
   \delta_2, \ldots, \delta_{N}}^{\alpha_1, \alpha_2, \ldots, \alpha_{N}}(t)
\end{equation}
is the desired projected correlation function. The correlation function
$C^{(N)}(t)$ is a tensor with antisymmetric groups of indices, hence only the
antisymmetric part of $\mathcal M$ contributes to the resulting correlation
function $C_{\mathcal M}(t)$. This antisymmetric part can be written as
\begin{equation}
 \mathcal M_{-}(\boldsymbol A^{(B_a)}\{\alpha\}, \boldsymbol
 A^{(B_b)}\{\alpha\}, \ldots, \boldsymbol A^{(B_a)}\{\delta\}, \boldsymbol
 A^{(B_b)}\{\delta\}, \ldots)
\end{equation}
and a modified list 
\begin{multline}
 (n_{B_a}!\,n_{B_b}!\ldots)^2 \, L_{\mathcal M}(\boldsymbol A^{(B_a)}\{\delta\},
  \boldsymbol A^{(B_b)}\{\delta\}, \ldots, \boldsymbol A^{(u)}\{\xi\},
  \boldsymbol A^{(d)}\{\xi\}, \boldsymbol A^{(s)}\{\xi\}) \\ =
  \sum_{\boldsymbol A^{(B_a)}\{\alpha\}, \boldsymbol A^{(B_b)}\{\alpha\},
    \ldots} L^{(N)}(\boldsymbol A^{(B_a)}\{\alpha\}, \boldsymbol
  A^{(B_b)}\{\alpha\}, \ldots, \boldsymbol A^{(u)}\{\xi\}, \boldsymbol
  A^{(d)}\{\xi\}, \boldsymbol A^{(s)}\{\xi\}) \\ \times \mathcal
  M_{-}(\boldsymbol A^{(B_a)}\{\alpha\}, \boldsymbol A^{(B_b)}\{\alpha\},
  \ldots, \boldsymbol A^{(B_a)}\{\delta\}, \boldsymbol A^{(B_b)}\{\delta\},
  \ldots)
\end{multline}
can be defined. Then this modified list can be used to calculate the
projected correlation function
\begin{multline}
C_{\mathcal M}(t) = \frac{1}{\mathcal N}\sum_{\boldsymbol A}
F^{(N)}_{-}(\boldsymbol A^{(B_a)}\{\delta\}, \ldots, \boldsymbol
A^{(u)}\{\xi\}, \ldots) \cdot L_{\mathcal M}(\boldsymbol A^{(B_a)}\{\delta\},
\ldots, \boldsymbol A^{(u)}\{\xi\},\ldots).
 \label{eqn:CM_final}
\end{multline}
Here the sum goes over all tuples of antisymmetric sets of indices.  Only
those components of $F^{(N)}_{-}$ contribute to the correlation function for
which the corresponding component of $L_{\mathcal M}^{(N)}$ is nonzero. The
number of contributing components is always smaller or equal to the number of
components that would be necessary to evaluate if the complete correlation
function were of interest. This fact can be exploited in the generation of the
lists of operations $\Lambda^{(1)}, \Lambda^{(2)}, \ldots, \Lambda^{(N)}$.

\section{Multiple sources/sinks\label{source_sink}}
\label{sec:many_sources}

In the previous sections the number of quark sources was set to one. In this
case due to the Pauli principle the maximal number of baryons is restricted in
such a way that only 12 quarks of each flavor are allowed in the system. This
restriction can be circumvented by introducing additional quark sources. When
we have $N_s$ mutually orthogonal quark sources, an additional source-index
$s$ ranging from $1$ to $N_s$ can be introduced to each quark operator. Then
the baryon operators at the source also have to be modified accordingly,
\begin{equation}
 \overline B^{\alpha;s} = \varepsilon^{abc} \, (\Gamma_1)^{\alpha\beta}
 (\overline q_1)^{\beta;a;s} \, [ (\overline q_2)^{\gamma;b;s}
   (\Gamma_2)^{\gamma\delta} (\overline q_3)^{\delta;c;s} ].
\end{equation}
As a consequence, the generation of the objects $G^B$ has to be modified in
such a way that the fact that all quarks of a baryon originate from the same
source is respected.\footnote{Although this restriction is not necessary for
  the algorithm described here, it is introduced to keep our notation simple
  and to allow for a cleaner presentation of the algorithm. The generalization
  to other cases is straightforward.} This can be achieved by the
straightforward modification: the indices $\xi^{(q_i)}$ are promoted to
combined spinor-color-source indices which range from $1$ to $12N_s$. Then the
modified $\tilde G^B$ becomes
\begin{equation}
 \tilde G^{B}(\alpha,s;\xi^{(q_1)}, \xi^{(q_2)}, \xi^{(q_3)}) := \delta^{s,
   s(\xi^{(q_1)})} \delta^{s, s(\xi^{(q_2)})} \delta^{s, s(\xi^{(q_3)})} \,
 G^{B}(\alpha;\kappa(\xi^{(q_1)}), \kappa (\xi^{(q_2)}),
 \kappa(\xi^{(q_3)})), \label{eqn:G_tilde}
\end{equation}
where $s(\xi)$ is the source part of the combined index $\xi$ and
$\kappa(\xi)$ is the spinor-color part of $\xi$.  The sink part also needs to
be modified by using different sink functions $s_1(\vec x), s_2(\vec x),
\ldots$ in the generation of the blocks $f_B^{q_1, q_2, q_3}(t, \delta;
\alpha, \beta, \gamma; a, b, c)$.

To simplify the notation combined spinor-source indices $\chi$ and $\psi$ can
be introduced to replace the former spinor indices of the baryons at the
source and at the sink, respectively.  Thus, the objects $\tilde G$ and the
modified blocks $\tilde f_B^{q_1, q_2, q_3}$ can be written as
\begin{align}
  \begin{split}
 \tilde G^{B}(\chi; \xi^{(q_1)}, \xi^{(q_2)}, \xi^{(q_3)}) &= \delta^{s(\chi),
   s(\xi^{(q_1)})} \delta^{s(\chi), s(\xi^{(q_2)})} \delta^{s(\chi),
   s(\xi^{(q_3)})} \\ &\qquad \cdot G^{B}(\alpha(\chi); \kappa(\xi^{(q_1)}),
 \kappa(\xi^{(q_2)}), \kappa(\xi^{(q_3)})),
 \end{split} \\
 \tilde f_B^{q_1, q_2, q_3}(t, \psi; \xi^{(q_1)}, \xi^{(q_2)}, \xi^{(q_3)}) &=
 \sum_{\vec x}s_{s(\psi)}(\vec x) \left< B_{\alpha(\psi)}(\vec x, t) \cdot
 q_1^{\xi^{(q_1)}} q_2^{\xi^{(q_2)}} q_3^{\xi^{(q_3)}} \right>.
\end{align}
Here $s(\chi)$ is the source part of the index $\chi$, $s(\psi)$ is the sink
part of the index $\psi$, and $\alpha(\chi)$ and $\alpha(\psi)$ are the spinor
parts of the indices $\chi$ and $\psi$, respectively.

The recursion relations (\ref{eqn:recr_L}) and (\ref{eqn:recr_F}) for $L$ and
$F_-$ remain valid with the only difference that $\tilde f_B^{q_1, q_2, q_3}$
and $\tilde G_B$ are used instead of $f_B^{q_1, q_2, q_3}$ and $G_B$. The
correlation function can be calculated similarly to (\ref{eqn:C_final}) by
evaluating the contraction
\begin{multline}
C^{(N)}(t; \boldsymbol A^{(B_a)}\{\psi\}, \ldots, \boldsymbol A^{(B_a)}\{\chi\},
\ldots) \\ = \frac{1}{\mathcal N} \sum_{\substack{\boldsymbol A^{(q_i)}\{\xi\}
    \\ i\in\{a,b,c\}}} F^{(N)}_{-}(\boldsymbol A^{(B_a)}\{\psi\}, \ldots,
\boldsymbol A^{(u)}\{\xi\}, \ldots) \cdot L^{(N)}(\boldsymbol
A^{(B_a)}\{\chi\}, \ldots, \boldsymbol A^{(u)}\{\xi\},\ldots).
 \label{eqn:C_final_moresource}
\end{multline}
Here the only difference is that the result $C^{(N)}(t; \boldsymbol
A^{(B_a)}\{\psi\}, \ldots, \boldsymbol A^{(B_a)}\{\chi\}, \ldots)$ instead of
only spin indices now possesses combined spinor-source indices.

The upper bounds for the computational effort for the construction of the
tensors $F_-$ can be generalized in a straightforward way to
\begin{equation}
 P(n^{(m)}_{B_1}, n^{(m)}_{B_2}, \ldots) = \prod_i C(n^{(m)}_{B_i},
 4N_s-n^{(m)}_{B_i}) \prod_j C(n^{(m)}_{q_j}, 12N_s-n^{(m)}_{q_j})
\end{equation}
and
\begin{multline}
 Q_{B_k}(n^{(m)}_{B_1}, n^{(m)}_{B_2}, \ldots) = 
\prod_i 
\begin{cases}
C(n^{(m)}_{B_i}, 0, 4N_s-n^{(m)}_{B_i}) & \text{for} \ i \neq k \\
C(n^{(m)}_{B_i}, 1, 4N_s-1-n^{(m)}_{B_i}) & \text{for} \ i = k
\end{cases} \\
\times \prod_j 
\begin{cases}
C(n^{(m)}_{q_j}, 0, 12N_s-n^{(m)}_{q_j}) & \text{for} \ j \neq k \\
C(n^{(m)}_{q_j}, N(q_j,B_k), 12N_s-n^{(m)}_{q_j}-N(q_j,B_k)) & \text{for} \ j = k.
\end{cases}
\end{multline}

\section{Atomic nuclei}
\label{sec:nuclei}

In this section the important special case of atomic nuclei, that is, systems
consisting of protons and neutrons is discussed.  For the calculation the
nucleon operators
\begin{subequations}
\begin{align}
 P_{\alpha} &= \varepsilon_{abc} \, (\Gamma_1)_{\alpha\beta} u_{\beta;a} \, [
   u_{\gamma;b} (\Gamma_2)_{\gamma\delta} d_{\delta;c} ], \\
 N_{\alpha} &= \varepsilon_{abc} \, (\Gamma_1)_{\alpha\beta} d_{\beta;a} \, [
   u_{\gamma;b} (\Gamma_2)_{\gamma\delta} d_{\delta;c} ]
\end{align}
are used at the sink and the operators
\begin{align}
 \overline P^{\alpha} &= \varepsilon^{abc} \, (\Gamma_1)^{\alpha\beta}
 \overline u^{\beta;a} \, [ \overline u^{\gamma;b} (\Gamma_2)^{\gamma\delta}
   \overline d^{\delta;c} ], \\
 \overline N^{\alpha} &= \varepsilon^{abc} \, (\Gamma_1)^{\alpha\beta}
 \overline d^{\beta;a} \, [ \overline u^{\gamma;b} (\Gamma_2)^{\gamma\delta}
   \overline d^{\delta;c} ]
\end{align}
are used at the source.
\end{subequations}
There are two common choices for the set of matrices $(\Gamma_1,
\Gamma_2)$. The first choice, $\Gamma_1 = \mathds 1$ and $\Gamma_2 = C\gamma_5$,
where $C$ is the charge conjugation matrix, gives fully relativistic nucleon
operators. In case of the second choice, $\Gamma_1 = P_{nr}$ and $\Gamma_2 =
C\gamma_5P_{nr}$, the projection $P_{nr}=(1+\gamma_4)/2$ to the
``nonrelativistic'' spinor components is inserted. In this case, if the Dirac
representation of the $\gamma$-matrices is used, only the upper two
spinor components contribute to the operators. Therefore, by using
nonrelativistic operators the computational effort can be reduced
significantly.

Introducing the variables $n_P$ and $n_N$ to denote the number of protons and
neutrons in the system, the recursion relations of adding one proton or
one neutron can be written as
\begin{subequations}
\begin{align}
 L^{(n_P+1, n_N)} &= L^{(n_P, n_N)} \bullet G_P,\label{eqn:rec_L_P}\\
 L^{(n_P, n_N+1)} &= L^{(n_P, n_N)} \bullet G_N,\label{eqn:rec_L_N}\\
 F_-^{(n_P+1, n_N)} &= F_-^{(n_P, n_N)} \bullet f_P^{u,u,d},\label{eqn:rec_F_P}\\
 F_-^{(n_P, n_N+1)} &= F_-^{(n_P, n_N)} \bullet f_N^{d,u,d},\label{eqn:rec_F_N}
\end{align}
\end{subequations}
with the starting conditions either
\begin{subequations}
\begin{equation}
L^{(0,1)} = G_N \qquad \text{and} \qquad F_-^{(0,1)} = f_N^{d,u,d}
\end{equation}
or
\begin{equation}
L^{(1,0)} = G_P\qquad \text{and} \qquad F_-^{(1,0)} = f_P^{u,u,d}.
\end{equation}
\end{subequations}

The upper bound for the number of components of $F_-$ at each stage is
\begin{multline}
 P(n_P, n_N) = C(n_P, D-n_P) C(n_N, D-n_N) \\ 
C(2n_P+n_N, 3D-2n_P-n_N) C(n_P+2n_N, 3D-n_P-2n_N) ,
\end{multline}
where $D$ denotes the effective number of spinor components. For relativistic
operators $D=4$ and for nonrelativistic operators $D=2$. The upper bounds for
the number of operations to add a proton or a neutron to $F_-^{(n_P, n_N)}$
are
\begin{subequations}
\begin{align}
 Q_P(n_P, n_N) &= C(n_P, 1, D-1-n_P) C(n_N, D-n_N) \notag C(2n_P+n_N, 2,
 3D-2-2n_P-n_N) \\ &\qquad C(n_P+2n_N, 1, 3D-1-n_P-2n_N), \\
 Q_N(n_P, n_N) &= C(n_P, D-n_P) C(n_N, 1, D-1-n_N) \notag C(2n_P+n_N, 1,
 3D-1-2n_P-n_N) \\ &\qquad C(n_P+2n_N, 2, 3D-2-n_P-2n_N).
\end{align}
\end{subequations}

\begin{figure}
 \begin{center}
  \includegraphics[width=0.5\textwidth]{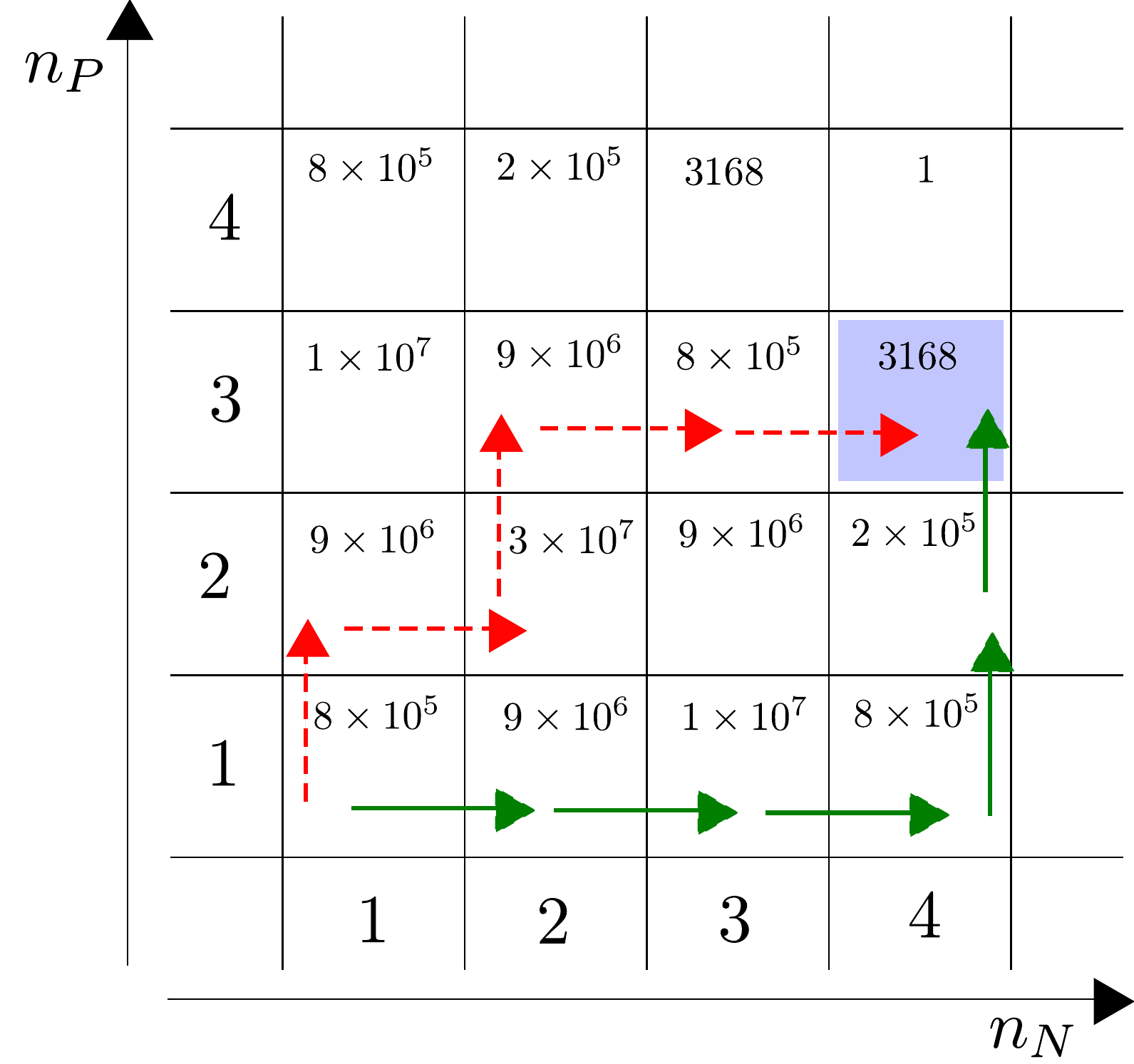}
  \caption{Two possible paths for the recursion relation to reach
    $F_-^{(3,4)}$ in the case of relativistic operators. The green (solid)
    path is more efficient than the red (dashed) path. The numbers in each box
    are the values of $P(n_P,n_N)$, that is, the upper bound for the number of
    components of $F_-^{(n_P,n_N)}$.\label{fig:opt_path}}
 \end{center}
\end{figure}

To calculate the tensor $F_-^{(N_P,N_N)}$ for fixed values of $N_P$ and $N_N$
one can choose several different orders of the recursion operations
(\ref{eqn:rec_L_P}--\ref{eqn:rec_F_N}). One could for example start with
adding only neutrons until $F_-^{(0,N_N)}$ is reached and then start to add
only protons until the final result $F_-^{(N_P,N_N)}$ is obtained. For a first
estimate of which order is best the values of the function $P(n_P, n_N)$ at
the intermediate stages can be used.  Figure \ref{fig:opt_path} gives these
values for relativistic operators and compares two possible ``paths'' leading
to the same state.
It can be seen that for a given $A=n_N+n_P$ the tensors with a minimal number
of components correspond to $A=n_N$ or $A=n_P$. After investigating $Q_P(n_P,
n_N)$ and $Q_N(n_P, n_N)$ it was found that the estimated operation count is
minimal when first all the baryons of one type, e.g. neutrons, are added
before adding any of the other type. Several different paths for several
different nuclei have been tried numerically and it was found that even with
the reduction due to the not required components these paths were still the
most efficient in all tested cases.

The lists of operations $\Lambda^{(n_P, n_N)}_P$ and $\Lambda^{(n_P, n_N)}_N$
for the addition of protons or neutrons to $F_-^{(n_P, n_N)}$ have been
constructed explicitly for a broad range of nuclei. These lists are different
for each choice of the numbers $N_P$ and $N_N$ even for the same intermediate
stages $(n_P, n_N)$ due to the fact that different components can be ignored
depending on $N_P$ and $N_N$.  In each case the path corresponding to adding
the $N_N$ neutrons before the addition of the $N_P$ protons was found to be the
fastest in all systems with $N_N > N_P$.

The generation of the lists takes less than an hour on a standard desktop
computer for all single nuclei accessible with one quark source and
relativistic operators. In the nonrelativistic case the generation of the
lists takes less than 0.1 seconds on the same computer.

The computational effort of generating the correlation functions of atomic
nuclei is dominated by the operations necessary to construct the tensors
$F_-$. All other computational tasks, such as the summation in equation
(\ref{eqn:C_final}) or the construction of $f_P^{u,u,d}$ and $f_N^{d,u,d}$ can
be neglected. Table \ref{tbl:num_ops1} shows the number of operations required
for the construction of atomic nuclei with one quark source and relativistic
operators. Here each operation is an element of a list $\Lambda$ required for
the given nucleus and amounts to a complex addition and multiplication. The
numbers are also compared to the na\"ive numbers of operations that would be
required if one evaluated all Wick contractions and spinor and color loops.
The same information for the case of nonrelativistic operators is shown in
Table \ref{tbl:num_ops2}.  A calculation with nonrelativistic operators and
two quark sources according to Section \ref{source_sink} was also
performed. The resulting numbers of operations are shown in Table
\ref{tbl:num_ops3}.

\begin{figure}
  \begin{center}
   \subfloat[$1.05 \cdot 10^9$ operations]{\includegraphics[width=0.3\textwidth]{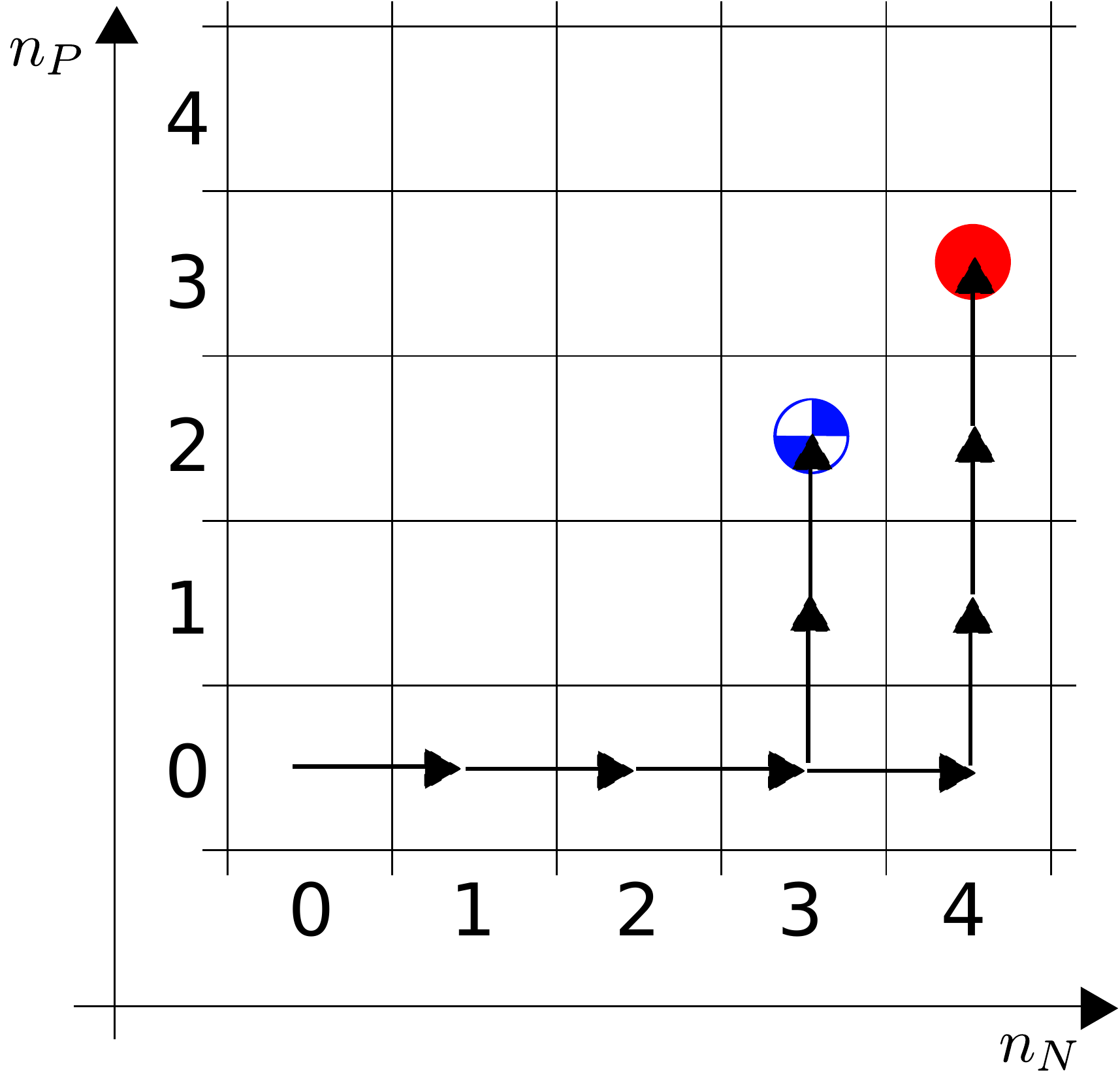}}
   \subfloat[$4.03 \cdot 10^9$ operations]{\includegraphics[width=0.3\textwidth]{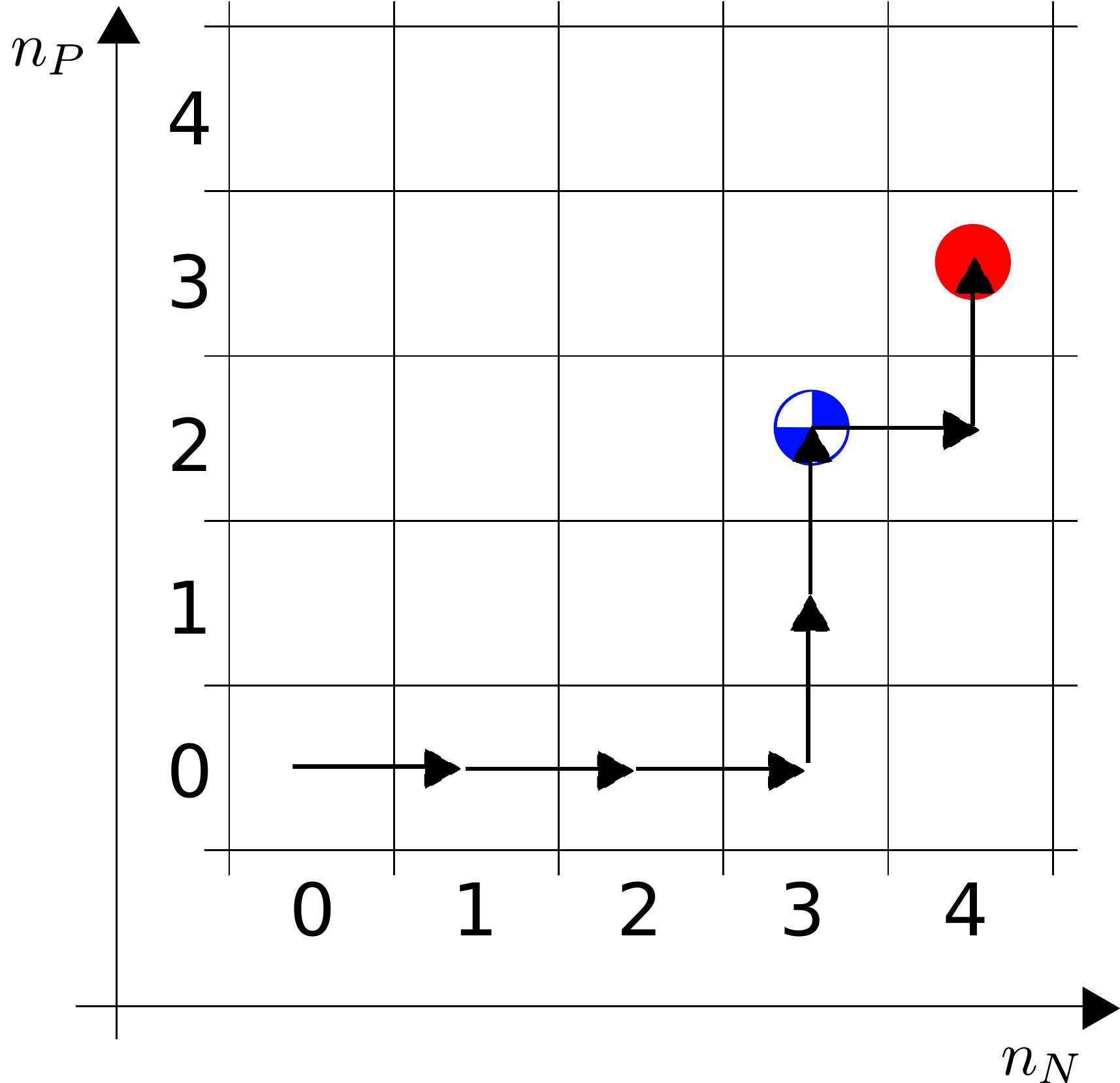}}
   \subfloat[$4.48 \cdot 10^9$ operations]{\includegraphics[width=0.3\textwidth]{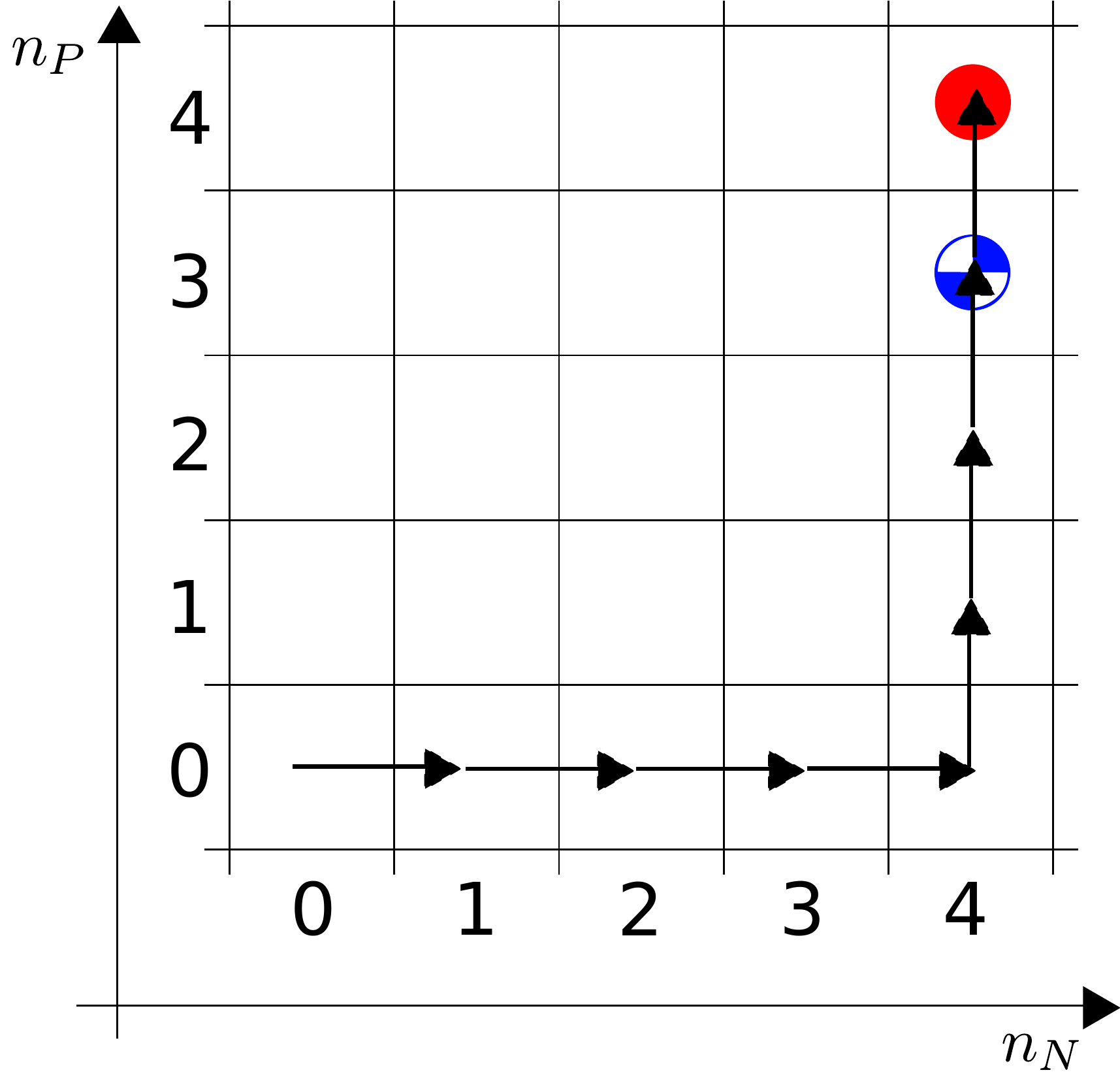}}
   \caption{Three representative cases for the combined calculation of two
     atomic nuclei. The red (filled) and blue (half filled) dots represent the
     nuclei that are to be calculated and the arrows indicate the order in
     which protons and neutrons are added. In the case (a) a speedup of about
     $10\%$ can be reached. In case (b) more operations are required than for
     the separate calculation of the indicated nuclei. In case (c) the blue
     (half filled) nucleus can be calculated without additional
     effort. Therefore in this case a speedup of about $47\%$ can be observed.
        \label{fig:three_paths}}
  \end{center}
\end{figure}

When only one specific nucleus is of interest, it is advantageous to add one
type of baryon after another in decreasing order of the total number of
baryons of the given species in the final correlation function.  Additional
smaller nuclei that lie on this path can then be calculated with a small
overhead depending on the desired spin states. In the general case, where
correlation functions of intermediate nuclei not lying on this path are of
interest, the situation is more complicated. Although the components that were
calculated for the largest nuclei can be reused for smaller nuclei, the
optimal path depends on the number, type and spin states of these nuclei in a
nontrivial way. Therefore, no general rule can be given for the best path,
but it should be decided on a case by case basis.  Figure
\ref{fig:three_paths} presents three representative cases for the calculation
of combinations of two nuclei with all possible spin states in the fully
relativistic case. For each case the number of required operations for the
recursive construction of $F_-$ is given.
When compared with Table \ref{tbl:num_ops1} these numbers show that calculating
several nuclei at once in a recursive manner can be effective, but care must be taken
to choose a good path for the recursive construction.

The numbers presented in this section show that the calculation of the
correlation function of atomic nuclei using the recursive approach discussed
in this paper is by many orders of magnitude more efficient than the na\"ive
computation and in certain cases can be even more effective when several
nuclei are to be calculated at once.

\section{Comparison of efficiency}
\label{sec:efficiency}

In this section we compare the efficiency of the procedure described in the
previous sections with the unified contraction algorithm introduced in
Ref.\ \cite{Doi:2012xd} and the determinant method introduced in
Ref.\ \cite{Detmold:2012eu}.

The method of Ref.\ \cite{Doi:2012xd} requires the construction of a unified
list of contractions. This list is identical to the tensor $L^{(N_P,N_N)}$
used in this paper except that it is not stored in an antisymmetrized
form. Hence the number of entries $N_\text{list}$ is by a factor
$n_u!\,n_d!\,n_s!$ larger than the number of entries in $L^{(N_P,N_N)}$. The
generation of the unified contraction list in Ref.\ \cite{Doi:2012xd} is done
by explicitly applying all possible permutations of quark indices. The effort
associated with these permutations scales with $n_u!\,n_d!\,n_s!$ and even for
systems of moderate sizes the generation of the list requires
supercomputers. In contrast to this our recursive construction of the similar
objects $L$ requires about a second on today's desktop computers for all nuclei
accessible with one quark source and relativistic operators.

Once the list is generated the number of evaluations of $f_B^{(q_1,q_2,q_3)}$
required on each gauge configuration is
$AN_\text{list}/2^A=AN_\text{contr}$. In our approach the number of
evaluations of $f_B^{(q_1,q_2,q_3)}$ is equal to the number of elements
$N_\Lambda$ in the operation list $\Lambda$ required for the
recursion. Therefore, the ratio of the efficiency of our algorithm and that of
the unified contraction algorithm can be roughly estimated as
$AN_\text{contr}/N_\Lambda$.

To be able to compare our method with the performance numbers listed in
Ref.\ \cite{Doi:2012xd} the different spin components have to be computed
separately.  We stress that this is less efficient than calculating all
components at once if one is interested in all components.  Individual spin
components can be computed using the technique described in Section
\ref{angular_momentum}.  If the spin indices of the desired
component are $(\gamma_1, \gamma_2, \ldots, \gamma_A)$ at the source and
$(\gamma_1', \gamma_2', \ldots, \gamma_A')$ at the sink then the projection tensor
\begin{equation}
 \mathcal M_{\delta_1\delta_2\ldots\delta_A}^{\alpha_1\alpha_2\ldots\alpha_A}
 =
 \delta^{\delta_1\gamma_1}\delta^{\delta_2\gamma_2}\ldots\delta^{\delta_A\gamma_A}
 \,
 \delta_{\alpha_1\gamma_1'}\delta_{\alpha_2\gamma_2'}\ldots\delta_{\alpha_A\gamma_A'}
\end{equation}
is to be used. Since only the choice for the spin component at the source
enters the calculation of the correlation function one can choose $\gamma_i =
\gamma_i'$ for the efficiency comparison.

Table \ref{tbl:eff_nonrel1} shows the comparison between our method and the
unified contraction algorithm in the nonrelativistic case. Table
\ref{tbl:eff_rel1} presents the same comparison in the case of
relativistic operators. In Table \ref{tbl:eff_nonrel2} the efficiency in
the case of nonrelativistic operators and two quark sources is shown.  For
the sake of readability not all components are tabulated in the case of
relativistic operators. More precisely only components with the minimal
possible number of lower half spinor indices are listed. This includes all
nonzero components calculated in Ref.\ \cite{Doi:2012xd}. The computational
effort associated with the components not listed is in all cases of the same
order of magnitude as the listed components with the same $N_P$ and $N_N$.

The method of Ref.\ \cite{Detmold:2012eu} uses determinants for the
calculation of the quark level permutations in the correlation function for a
fixed structure of color/spinor indices and spatial location of the operators
at the source and at the sink. This algorithm scales as
\begin{equation}
 n_u^3  n_d^3  n_s^3 \cdot N_w N_w',
\end{equation}
where $N_w$ and $N_w'$ are the number of such independent structures up to
permutations of quarks at the source and at the sink, respectively.  This
algorithm is especially efficient in the case where the numbers of quarks
$n_u$, $n_d$ and $n_s$ are chosen such that all possible spinor and
color DoFs are fully saturated both at the source and at the sink.  In this
case $N_w = N_w' = 1$. Such combinations of operators can be found for the
nuclei $^4\text{He}$, $^8\text{Be}$, $^{12}\text{C}$, $^{16}\text{O}$ and
$^{28}\text{Si}$ for which concrete results are presented in
Ref.\ \cite{Detmold:2012eu}.  However, in the general case the numbers $N_w$
and $N_w'$ may become very large and in fact scale exponentially as already
noted in Ref.\ \cite{Detmold:2012eu}.

If the individual baryons are to have a complex spatial structure, which is
required e.g.\ for the projection to a definite momentum and angular momentum,
it is very difficult to find a combination of operators for which $N_w$ and
$N_w'$ remain small. In such cases the algorithm presented in this paper,
which can perform the construction of baryon blocks with complex spatial
structure in advance of the calculation, can be more advantageous.

\section{Summary}
\label{sec:conclusions}

We introduced a procedure to compute the correlation functions of multibaryon
systems using a recursive technique. In a first step a list of required
components is generated in a recursive manner, where several antisymmetry
properties of the list are exploited. In a second step a product of blocks of
quark propagators is constructed recursively on each gauge
configuration. During the construction both the antisymmetry of this product
and the reduction due to the sparse nature of the list of required components
are exploited at each intermediate step.

The individual baryons in the system can be projected to any momentum state
prior to the calculation, e.g. to zero momentum to extract the ground state
mass. Different quark sources and baryon sinks can be used to create
correlation functions with spatial structure and an arbitrary number of
baryons in the system. The procedure can be employed for a broad range of
multibaryon systems. Systems with quantum numbers of atomic nuclei were
discussed in detail.

Our technique was compared in detail with the na\"ive method and the method proposed in
\cite{Doi:2012xd} and it was found that a significant speedup in all cases
with $A>2$ is possible. For the construction of the $^4\text{He}$ and
$^8\text{Be}$ correlation functions with relativistic operators $\mathcal
O(10^8)$ operations are required. In the nonrelativistic approximation for
$^4\text{He}$ the required number of operations is only $\mathcal O(10^4)$.

\section*{Acknowledgements}
\label{sec:acknowledgements}

We would like to thank Zolt\'an Fodor for the stimulating suggestions and
continuous support throughout the project. This work is supported in part by
the DFG Grant No.\ SFB/TR55 and the GSI Grant No.\ WFODOR1012.

\bibliography{multi_baryon_recursion}
\bibliographystyle{JHEP}


\clearpage
\renewcommand{\sectionmark}[1]{\markright{#1}{}}
\sectionmark{Tables}

\begin{table}
 \caption{Number of operations (each operation is a complex multiplication and
   addition) required to compute all independent spinor components of the
   correlation function with $N_P$ protons and $N_N$ neutrons with \emph{one
     quark source} and \emph{relativistic operators}. Both the na\"ive number
   and the number using the recursive approach are given. $\eta$ is the gain
   factor, that is, the ratio of the na\"ive and the recursive numbers of
   operations.\label{tbl:num_ops1}}
 \begin{center}
 \begin{tabular}{lllll}
  $N_P$ & $N_N$ & No. of op. & Na\"ive no. of op. & $\eta$\\
  \hline
  0 & 2 & 199584 & 15925248 & 79.8 \\
  0 & 3 & 5825088 & $8.3 \times 10^{11}$ & $1.4 \times 10^5$ \\
  0 & 4 & 54768672 & $1.1 \times 10^{17}$ & $1.9 \times 10^9$ \\
  1 & 1 & 474048 & 11943936 & 25.2 \\
  1 & 2 & 19241280 & $5.5 \times 10^{11}$ & $2.9 \times 10^4$ \\
  1 & 3 & 109789200 & $6.7 \times 10^{16}$ & $6.1 \times 10^8$ \\
  1 & 4 & 179769600 & $1.7 \times 10^{22}$ & $9.2 \times 10^{13}$ \\
  2 & 2 & 531321120 & $5.7 \times 10^{16}$ & $1.1 \times 10^8$ \\
  2 & 3 & 756897264 & $1.3 \times 10^{22}$ & $1.7 \times 10^{13}$ \\
  2 & 4 & 291957888 & $5.3 \times 10^{27}$ & $1.8 \times 10^{19}$ \\
  3 & 3 & 2905079520 & $4.9 \times 10^{27}$ & $1.7 \times 10^{18}$ \\
  3 & 4 & 404946240 & $3.0 \times 10^{33}$ & $7.5 \times 10^{24}$ \\
  4 & 4 & 448496928 & $2.8 \times 10^{39}$ & $6.2 \times 10^{30}$
 \end{tabular}
 \end{center}
\end{table}

\begin{table}
 \caption{Number of operations (each operation is a complex multiplication and
   addition) required to compute all independent spinor components of the
   correlation function with $N_P$ protons and $N_N$ neutrons with \emph{one
     quark source} and \emph{nonrelativistic operators}. Both the na\"ive number
   and the number using the recursive approach are given. $\eta$ is the gain
   factor, that is, the ratio of the na\"ive and the recursive numbers of
   operations.\label{tbl:num_ops2}}
 \begin{center}
 \begin{tabular}{lllll}
  $N_P$ & $N_N$ & No. of op. & Na\"ive no. of op. & $\eta$\\
  \hline
  0 & 2 & 504 & 995328 & 1974.9 \\
  1 & 1 & 2664 & 746496 & 280.2 \\
  1 & 2 & 6048 & $8.6\times 10^9$ & $1.4\times 10^6$ \\
  2 & 2 & 10980 & $2.2\times 10^{14}$ & $2.0\times 10^{10}$
 \end{tabular}
 \end{center}
\end{table}

\begin{table}
 \caption{Number of operations (each operation is a complex multiplication and
   addition) required to compute all independent spinor components of the
   correlation function with $N_P$ protons and $N_N$ neutrons with \emph{two
     quark sources} and \emph{nonrelativistic operators}. Both the na\"ive number
   and the number using the recursive approach are given. $\eta$ is the gain
   factor, that is, the ratio of the na\"ive and the recursive numbers of
   operations.\label{tbl:num_ops3}}
 \begin{center}
 \begin{tabular}{lllll}
  $N_P$ & $N_N$ & No. of op. & Na\"ive no. of op. & $\eta$\\
  \hline
0 & 2 & 3024 & 995328 & 329.1 \\
0 & 3 & 1052136 & $1.3 \times 10^{10}$ & $1.2 \times 10^4$ \\
0 & 4 & 18881568 & $4.2 \times 10^{14}$ & $2.2 \times 10^7$ \\
1 & 1 & 10656 & 746496 & 70.1 \\
1 & 2 & 42768 & $8.6 \times 10^9$ & $2.0 \times 10^5$ \\
1 & 3 & 6329016 & $2.6 \times 10^{14}$ & $4.1 \times 10^7$ \\
1 & 4 & 67720680 & $1.6 \times 10^{19}$ & $2.4 \times 10^{11}$ \\
2 & 2 & 103680 & $2.2 \times 10^{14}$ & $2.1 \times 10^9$ \\
2 & 3 & 10038672 & $1.3 \times 10^{19}$ & $1.3  \times 10^{12}$ \\
2 & 4 & 81850128 & $1.3 \times 10^{24}$ & $1.6 \times 10^{16}$ \\
3 & 3 & 338263368 & $1.3 \times 10^{24}$ & $3.5 \times 10^{15}$ \\
3 & 4 & 287427384 & $1.9 \times 10^{29}$ & $6.5 \times 10^{20}$ \\
4 & 4 & 448496928 & $4.2 \times 10^{34}$ & $9.5 \times 10^{25}$
 \end{tabular}
 \end{center}
\end{table}

\begin{table}
\caption{The efficiency of the presented algorithm for the calculation of
  individual spin components with \emph{nonrelativistic
    operators}. $N_\Lambda$ is the number of operations (complex
  multiplications and additions) required for the construction of $F_-$. $N_L$
  is the number of independent components of the tensor $L$. $N_\text{list}$
  and $N_\text{contr}$ are the number of entries in the unified contraction
  list and the number of independent contractions from
  Ref.\ \cite{Doi:2012xd}, respectively. (Numbers that are not presented in
  \cite{Doi:2012xd} are from our own calculations.) $AN_\text{contr.}/N_\Lambda$
  is approximately the ratio of the number of operations required for the
  unified contraction algorithm and for the algorithm presented in this paper
  if additions are taken as much faster than
  multiplications. \label{tbl:eff_nonrel1} }
\begin{center}
\begin{tabular}{llllllll}
$N_P$ & $N_N$ & Spin state & $N_\Lambda$ & $N_L$ & $N_\text{list}$ & $N_\text{contr}$ & $AN_\text{contr.}/N_\Lambda$ \\
\hline
0 & 2 & (0,1) & 504 & 21 & 1008 & 252 & 1 \\
1 & 1 & (0,0) & 189 & 21 & 756 & 189 & 2 \\
1 & 1 & (1,0) & 252 & 28 & 1008 & 252 & 2 \\
1 & 1 & (0,1) & 252 & 28 & 1008 & 252 & 2 \\
1 & 1 & (1,1) & 189 & 21 & 756 & 189 & 2 \\
1 & 2 & (0,0,1) & 4662 & 9 & 25920 & 3240 & 2.1 \\
1 & 2 & (1,0,1) & 4662 & 9 & 25920 & 3240 & 2.1 \\
2 & 2 & (0,1,0,1) & 10980 & 1 & 518400 & 32400 & 11.8
\end{tabular}
\end{center}
\end{table}

\begin{landscape}
\begin{longtable}{llllllll}
\caption{The efficiency of the presented algorithm for the calculation of
  individual spin components with \emph{relativistic operators}. $N_\Lambda$
  is the number of operations (complex multiplications and additions) required
  for the construction of $F_-$. $N_L$ is the number of independent components
  of the tensor $L$. $N_\text{list}$ and $N_\text{contr}$ are the number of
  entries in the unified contraction list and the number of independent
  contractions from Ref.\ \cite{Doi:2012xd}, respectively. (Numbers that are
  not presented in \cite{Doi:2012xd} are from our own
  calculations.) $AN_\text{contr.}/N_\Lambda$ is approximately the ratio of
  the number of operations required for the unified contraction algorithm and
  for the algorithm presented in this paper if additions are taken as much
  faster than multiplications.}\\
 $N_P$ & $N_N$ & Spin state & $N_\Lambda$ & $N_L$ & $N_\text{list}$ & $N_\text{contr}$ & $AN_\text{contr.}/N_\Lambda$ \\
 \hline\endhead \label{tbl:eff_rel1}
0 & 2 & (0,1) & 5544 & 231 & 11088 & 2772 & 1 \\
0 & 3 & (0,1,2) & 1360098 & 1110 & 4795200 & 599400 & 1.3 \\
0 & 3 & (0,1,3) & 1360098 & 1110 & 4795200 & 599400 & 1.3 \\
0 & 4 & (0,1,2,3) & 54768672 & 1845 & 1785369600 & 111585600 & 8.1 \\
1 & 1 & (0,0) & 2079 & 231 & 8316 & 2079 & 2 \\
1 & 1 & (1,0) & 2358 & 262 & 9432 & 2358 & 2 \\
1 & 1 & (0,1) & 2358 & 262 & 9432 & 2358 & 2 \\
1 & 1 & (1,1) & 2079 & 231 & 8316 & 2079 & 2 \\
1 & 2 & (0,0,1) & 381978 & 1311 & 3775680 & 471960 & 3.7 \\
1 & 2 & (1,0,1) & 381978 & 1311 & 3775680 & 471960 & 3.7 \\
1 & 3 & (0,0,1,2) & 11717937 & 2232 & 1349913600 & 84369600 & 28.8 \\
1 & 3 & (1,0,1,2) & 11717937 & 2232 & 1349913600 & 84369600 & 28.8 \\
1 & 3 & (0,0,1,3) & 11717937 & 2232 & 1349913600 & 84369600 & 28.8 \\
1 & 3 & (1,0,1,3) & 11717937 & 2232 & 1349913600 & 84369600 & 28.8 \\
1 & 4 & (0,0,1,2,3) & 141103602 & 1110 & 290013696000 & 9062928000 & 321.1 \\
1 & 4 & (1,0,1,2,3) & 141103602 & 1110 & 290013696000 & 9062928000 & 321.1 \\
2 & 2 & (0,1,0,1) & 8541864 & 2716 & 1407974400 & 87998400 & 41.2 \\
2 & 3 & (0,1,0,1,2) & 44343561 & 1311 & 266411980800 & 8325374400 & 938.7 \\
2 & 3 & (0,1,0,1,3) & 44343561 & 1311 & 266411980800 & 8325374400 & 938.7 \\
2 & 4 & (0,1,0,1,2,3) & 214572144 & 231 & 33798352896000 & 528099264000 & 14767 \\
3 & 3 & (0,1,2,0,1,2) & 163007703 & 231 & 30418517606400 & 475289337600 & 17494.5 \\
3 & 3 & (0,1,3,0,1,2) & 181280493 & 262 & 34500656332800 & 539072755200 & 17842.2 \\
3 & 3 & (0,1,2,0,1,3) & 181280493 & 262 & 34500656332800 & 539072755200 & 17842.2 \\
3 & 3 & (0,1,3,0,1,3) & 163007703 & 231 & 30418517606400 & 475289337600 & 17494.5 \\
3 & 4 & (0,1,2,0,1,2,3) & 293717796 & 21 & 3041851760640000 & 23764466880000 & 566364 \\
3 & 4 & (0,1,3,0,1,2,3) & 293717796 & 21 & 3041851760640000 & 23764466880000 & 566364 \\
4 & 4 & (0,1,2,3,0,1,2,3) & 448496928 & 1 & 229442532802560000 & 896259893760000 & $1.6\times 10^{7}$
\end{longtable}

\begin{longtable}{llllllll}
\caption{The efficiency of the presented algorithm for the calculation of
  individual spin components with \emph{nonrelativistic operators using two
    quark sources}. $N_\Lambda$ is the number of operations (complex
  multiplications and additions) required for the construction of $F_-$. $N_L$
  is the number of independent components of the tensor $L$. $N_\text{list}$
  and $N_\text{contr}$ are the number of entries in the unified contraction
  list and the number of independent contractions from
  Ref.\ \cite{Doi:2012xd}, respectively. (Numbers that are not presented in
  \cite{Doi:2012xd} are from our own calculations.) $AN_\text{contr.}/N_\Lambda$
  is approximately the ratio of the number of operations required for the
  unified contraction algorithm and for the algorithm presented in this paper
  if additions are taken as much faster than multiplications. The combined
  spinor-source indices are of the form $2 \alpha + s$ where $\alpha$ is the
  spinor part and $s$ is the source part.}\\
 $N_P$ & $N_N$ & spinor-source indices & $N_\Lambda$ & $N_L$ & $N_\text{list}$ & $N_\text{contr}$ & $AN_\text{contr.}/N_\Lambda$ \\
 \hline\endhead \label{tbl:eff_nonrel2}
0 & 2 & (0,2) & 504 & 21 & 1008 & 252 & 1 \\
0 & 3 & (0,1,2) & 330291 & 189 & 816480 & 102060 & 0.9 \\
0 & 3 & (0,2,3) & 330291 & 189 & 816480 & 102060 & 0.9 \\
0 & 4 & (0,1,2,3) & 18881568 & 441 & 426746880 & 26671680 & 5.7 \\
1 & 1 & (0,0) & 189 & 21 & 756 & 189 & 2 \\
1 & 1 & (2,0) & 252 & 28 & 1008 & 252 & 2 \\
1 & 1 & (0,2) & 252 & 28 & 1008 & 252 & 2 \\
1 & 1 & (2,2) & 189 & 21 & 756 & 189 & 2 \\
1 & 2 & (0,0,2) & 4662 & 9 & 25920 & 3240 & 2.1 \\
1 & 2 & (2,0,2) & 4662 & 9 & 25920 & 3240 & 2.1 \\
1 & 3 & (0,0,1,2) & 1100034 & 81 & 48988800 & 3061800 & 11.1 \\
1 & 3 & (2,0,1,2) & 1100034 & 81 & 48988800 & 3061800 & 11.1 \\
1 & 3 & (0,0,2,3) & 1100034 & 81 & 48988800 & 3061800 & 11.1 \\
1 & 3 & (2,0,2,3) & 1100034 & 81 & 48988800 & 3061800 & 11.1 \\
1 & 4 & (0,0,1,2,3) & 59747247 & 189 & 49380710400 & 1543147200 & 129.1 \\
1 & 4 & (2,0,1,2,3) & 59747247 & 189 & 49380710400 & 1543147200 & 129.1 \\
2 & 2 & (0,2,0,2) & 10980 & 1 & 518400 & 32400 & 11.8 \\
2 & 3 & (0,2,0,1,2) & 1717569 & 9 & 1828915200 & 57153600 & 166.4 \\
2 & 3 & (0,2,0,2,3) & 1717569 & 9 & 1828915200 & 57153600 & 166.4 \\
2 & 4 & (0,2,0,1,2,3) & 80357088 & 21 & 3072577536000 & 48009024000 & 3584.7 \\
3 & 3 & (0,1,2,0,1,2) & 31373721 & 21 & 2765319782400 & 43208121600 & 8263.3 \\
3 & 3 & (0,2,3,0,1,2) & 40214061 & 28 & 3687093043200 & 57610828800 & 8595.6 \\
3 & 3 & (0,1,2,0,2,3) & 40214061 & 28 & 3687093043200 & 57610828800 & 8595.6 \\
3 & 3 & (0,2,3,0,2,3) & 31373721 & 21 & 2765319782400 & 43208121600 & 8263.3 \\
3 & 4 & (0,1,2,0,1,2,3) & 225681807 & 9 & 1303650754560000 & 10184771520000 & 315902 \\
3 & 4 & (0,2,3,0,1,2,3) & 225681807 & 9 & 1303650754560000 & 10184771520000 & 315902 \\
4 & 4 & (0,1,2,3,0,1,2,3) & 448496928 & 1 & 229442532802560000 & 896259893760000 & $1.6\times 10^{7}$
\end{longtable}
\end{landscape}

\end{document}